# Micro-Ring Modulator Linearity Enhancement for Analog and Digital Optical Links

Sumilak Chaudhury, Karl Johnson, Chengkuan Gao, Bill Lin, *Senior Member, IEEE*, Yeshaiahu Fainman, *Fellow, IEEE*, and Tzu-Chien Hsueh, *Senior Member, IEEE*

***Abstract* — An energy/area-efficient low-cost broadband linearity enhancement technique for electro-optic micro-ring modulators (MRM) is proposed to achieve 6.1-dB dynamic linearity improvement in spurious-free-dynamic-range (ΔSFDR) with intermodulation distortions (IMD) and 17.9-dB static linearity improvement in integral nonlinearity (ΔINL) over a conventional notch-filter MRM within a 4.8-dB extinction-ratio (ER) full-scale range based on rapid silicon-photonics fabrication results for the emerging applications of various analog and digital optical communication systems.**

## I. Introduction

ANALOG and digital optical communication systems have become increasingly important in the emerging applications of radio-frequency (RF) photonic links [1], radio-over-fiber, antenna remoting, subcarrier transmission, phased array antenna control [2], high-speed Ethernet, and data-center interconnects [3]. The nonlinearity of electro-optic modulators used in the electrical-to-optical (E/O) interfaces induces data distortions, limits signal dynamic ranges, and dominates overall link performance, particularly in these broadband optical communication systems and applications. Many research projects have been focusing on mitigating the nonlinearity issues of Mach-Zehnder modulators (MZM) covering from high-linearity Lithium Niobate ($LiNbO_3$) discrete MZMs [4] to today's silicon ring-assist MZMs [5] and heterogeneously integrated III-V/silicon MZMs [6], which have reached higher than 110-dB·Hz$^{2/3}$ spurious-free-dynamic-range with intermodulation distortions (i.e., $SFDR_{IMD}$). However, even in the most advanced monolithic silicon-photonics (M-SiPh) process technology at GlobalFoundries (GF45SPCLO) [7], the on-chip area of an MZM is still in the range of multiple mm$^2$, this leads to the fact that an analog linear amplifier or a digital pulse-amplitude driver interfacing with an MZM has to treat the MZM electrode as a transmission line (TL) with the design effort of 50-Ω characteristic impedance and the power/area overhead of on-chip terminations as well as matching network [8], [9] when the RF signal frequencies of the analog optical links or the Nyquist frequencies of digital data links are close to or higher than 10 GHz, which renders sub-optimal on-chip energy/area efficiency and further induces the power density violations as well as device reliability issues.

Micro-ring modulators (MRM) have been widely used in high-speed pulse-amplitude modulation (PAM) digital data communications because their compact silicon areas and footprints, which can be more than 800× smaller than that of an MZM in GF45SPCLO, tremendously reduce the electrical signal traveling distances between the amplifier/driver outputs and MRM electrodes. In other words, an MRM-based E/O interface can be treated as lumped and small capacitive circuitry to enable twofold improvements in energy, area, and integration efficiencies. However, using optical power intensity modulation (IM) through the notch-filter-like power-gain response of an MRM (i.e., a notch-filter MRM) inevitably produces a very nonlinear E/O-conversion transfer function even across a small extinction-ratio (ER) full-scale (FS) range. This is why the usages of MRMs are confined within the applications of high-speed but low-resolution (i.e., 2 bits/symbol or PAM-4) digital data communications.

To retain the excellent energy/area efficiency and further extend the applications of MRMs, this paper proposes a low-cost linearity enhancement technique, without sophisticated digital linearization equalizers [3], to boost the $SFDR_{IMD}$ and integral nonlinearity (INL) of a dual-MRM based transmitter (TX) by 6.1 dB and 17.9 dB, respectively, over a conventional notch-filter-MRM based TX within a 4.8-dB ER FS range for both analog and digital optical communication systems.

## II. MRM Linearity Enhancement Technique

The proposed MRM linearity enhancement technique illustrated in Fig. 1(a) is established on top of a conventional notch-filter MRM ($MRM_1$) with the circuit configuration shown in the top half of Fig. 1(a) used for optical power IM-based E/O conversions. Therefore, the technical elaboration and The authors are with the Department of Electrical and Computer Engineering, University of California, San Diego, La Jolla, CA 92093 USA (e-mail: tzhsueh@ucsd.edu).

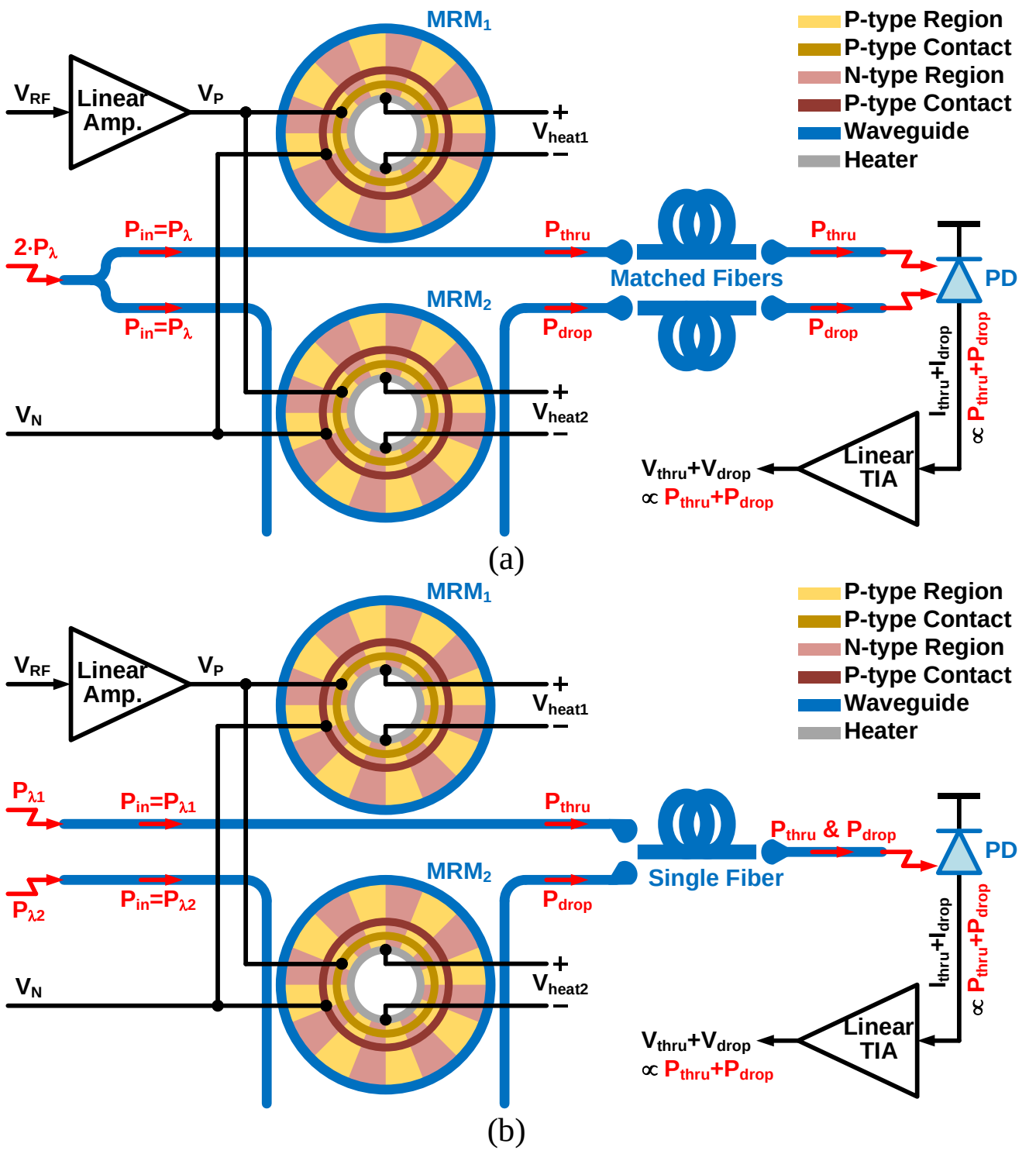

Fig. 1. The circuit schematic of the MRM linearity enhancement technique (Dual-MRMs: $MRM_1$ + $MRM_2$) on top of the conventional notch-filter MRM ($MRM_1$) with (a) a single laser wavelength and two matched fibers or (b) two laser wavelengths and a single fiber.

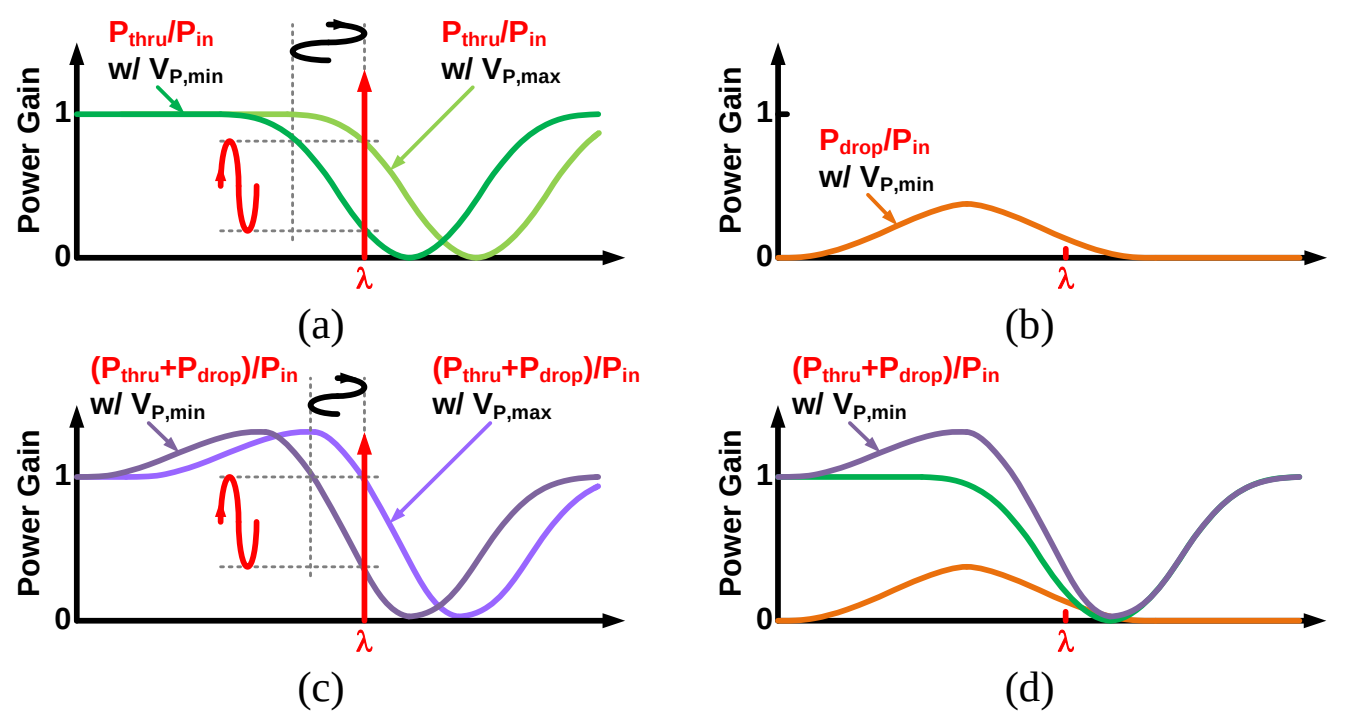

Fig. 2. The power-gain responses of (a) the notch-filter MRM ($MRM_1$) with the electrical signal $V_P$ modulation, (b) the bandpass-filter MRM ($MRM_2$), (c) Dual-MRMs with electrical signal $V_P$ modulation, and (d) Dual-MRMs achieved by summing the power responses of $MRM_1$ and $MRM_2$.

the performance comparisons in this paper should start with and refer to the case of this notch-filter MRM, whose E/O conversion can be represented by the signal processing flow from the electrical signal $V_P$ driven by the linear amplifier to the $MRM_1$ Through-Port output optical power $P_{thru}$ or power gain $P_{thru}/P_{in}$; note that $P_{in}$ (= $P_\lambda$) is a constant laser power injected into the $MRM_1$ Input-Port with a wavelength λ (λ-laser), not the electrical signal power at $V_P$. According to certain design specifications of an E/O interface and integrated photonic process technology, the free spectral range (FSR), radius, coupling coefficient, quality factor (Q), optical frequency band, and initial resonant frequency of $MRM_1$ can be properly determined, and then, by influencing the refractive and group indexes in the resonant cavity of $MRM_1$ during signal transmissions, the static control-knob of shifting the power-gain response w.r.t the wavelength axis is the thermal control (or heater) voltage $V_{heat1}$ while the dynamic (or high-speed) control-knob of shifting the power-gain response w.r.t the wavelength axis is the reverse bias ($V_N - V_P$) of the segmented PN junctions underneath the ring waveguide of $MRM_1$. As the $P_{thru}/P_{in}$ response shown in Fig. 2(a), the default spectrum and notch point of $P_{thru}/P_{in}$ can be designated by $V_{heat1}$ as the green curve when $V_P$ is set to its minimum ($V_{P,min}$), and in principle, the whole $P_{thru}/P_{in}$ response can be shifted to the position of the light-green curve when $V_P$ is set to its maximum ($V_{P,max}$). Therefore, if $V_P$ is continuously varying between $V_{P,min}$ and $V_{P,max}$, i.e., electrical signal FS, like the black sinusoid waveform in Fig. 2(a), then the optical power $P_{thru}$ (or power gain $P_{thru}/P_{in}$) at the $MRM_1$ Through-Port shall vary accordingly within the output FS like the red sinusoid waveform in Fig. 2(a) as well, which is the essential concept of the IM-based E/O conversion through a notch-filter MRM. Note that since $V_N$ is a constant voltage from a low-impedance voltage supply, the electrical bandwidth dominating the whole E/O interface is determined by the linear amplifier output impedance and PN junction capacitance of $MRM_1$ at $V_P$. Because of the compact footprint of an MRM in general and advanced M-SiPh process technology [7], the electrical bandwidth at $V_P$ can reasonably reach higher than 20 GHz without energy/area overhead due to the requirements of TLs and impedance matching network.

The mechanism of an E/O conversion through the notch-filter MRM described in Fig. 2(a) has two major sources of nonlinearity: The first one is how linear or straight the $P_{thru}/P_{in}$ response (green curve) can perform within the optical power FS (between the two horizontal dash lines), which can be quantified by the metric of INL for static and continuous signal-conversion transfer curves in general. The second one is how the $P_{thru}/P_{in}$ response can be linearly shifted within the electrical signal FS (between the green and light-green curves or between the two vertical dash lines), which can be covered by the metrics of $SFDR_{IMD}$ in the third intercept point ($IP_3$) measurements.

The proposed linearity enhancement technique is achieved by adding a bandpass-filter MRM ($MRM_2$), which is also driven by the same reverse bias of $MRM_1$ but has an independent thermal control voltage $V_{heat2}$, as shown in the bottom half of Fig. 1(a). The fundamentals of a notch-filer MRM and a bandpass-filter MRM are basically identical; their primary difference is that the input/output port configuration of $MRM_2$ renders its E/O conversion to be the signal processing flow from the electrical signal $V_P$ driven by the linear amplifier to the $MRM_2$ Drop-Port output optical power $P_{drop}$ or power gain $P_{drop}/P_{in}$, which behaves as a peaking response (orange curve) as shown in Fig. 2(b). Therefore, the key idea of this technique is to exploit the complementary power-gain responses of $MRM_1$ and $MRM_2$ to form a dual-MRM architecture (Dual-MRMs) as the full picture shown in Fig. 1(a) for overall linearity enhancement. This simple architecture contains multiple circuit and signal design parameters between $MRM_1$ and $MRM_2$ for optimizing the degree of the linearity enhancement, including the ring radii, coupling coefficients, laser injection powers, laser wavelengths, and thermal control voltages. The following technical description assumes that $MRM_1$ and $MRM_2$ have all identical design parameters, except $V_{heat1}$ and $V_{heat2}$ can independently set up the resonant frequencies of $MRM_1$ and $MRM_2$ to properly misalign the default notch point and peaking point of $P_{thru}/P_{in}$ (green curve) and $P_{drop}/P_{in}$ (orange curve), respectively, as shown in Fig. 2(d), when $V_P$ is set to $V_{P,min}$.

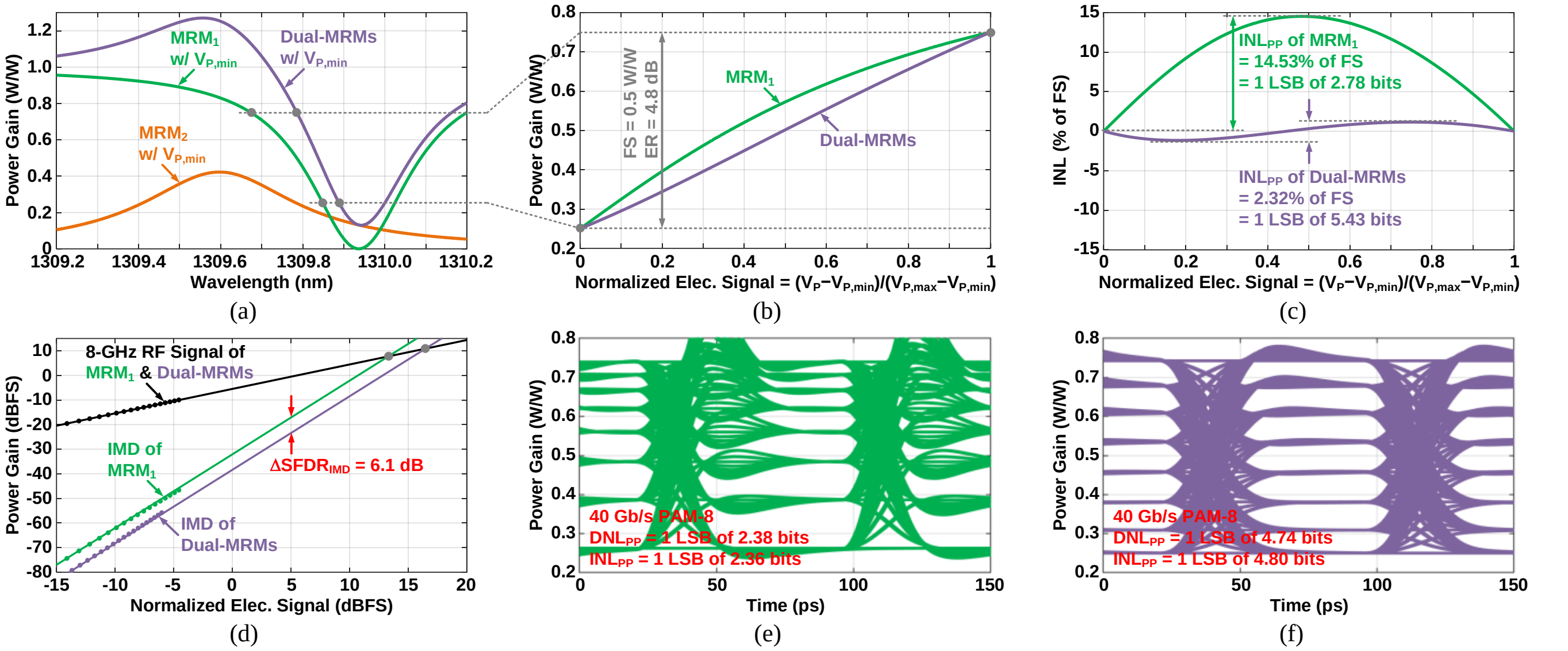

Fig. 3. The simulation results by using O-band MRMs in GF45SPCLO: (a) The power-gain responses of $MRM_1$, $MRM_2$, and Dual-MRMs when $V_P = V_{P,min}$. (b) The static E/O-conversion transfer curves of $MRM_1$ and Dual-MRMs. (c) The INLs of the static E/O-conversion transfer curves in (b). (d) The IMDs in the $IP_3$ measurements of $MRM_1$ and Dual-MRMs when $V_P$ is an 8-GHz RF signal, (e) The time-domain eye-diagram of the $MRM_1$ power gain when $V_P$ is 40-Gb/s PAM-8 random data. (f) The time-domain eye-diagram of the Dual-MRMs power gain when $V_P$ is 40-Gb/s PAM-8 random data.

Under the settings described above, if these two power-gain responses can be linearly and incoherently summed, which is further discussed at the end of this paper, the peaking spectrum of $P_{drop}/P_{in}$ can be used to compensate for the flat and non-steep spectrum of the nonlinear $P_{thru}/P_{in}$ region while the low power-gain and almost zero spectrum of $P_{drop}/P_{in}$ can negligibly affect the steep spectrum of the linear $P_{thru}/P_{in}$ region. Equivalently, the power-gain response of Dual-MRMs can be expressed as the $(P_{thru} + P_{drop})/P_{in}$ response illustrated by the purple curve in Fig. 2(d), whose linear region is pronouncedly larger than that of the $P_{thru}/P_{in}$ response. Once the reverse biases of $MRM_1$ and $MRM_2$ are simultaneously modulated by the electrical signal $V_P$ like the black sinusoid waveforms in Fig. 2(c), both $P_{thru}/P_{in}$ and $P_{drop}/P_{in}$ responses as well as their equivalent summation (i.e., the $(P_{thru} + P_{drop})/P_{in}$ response) can be all shifted together w.r.t the wavelength axis between the purple and light-purple curves. Because of the larger linear region and steeper slope of the $(P_{thru} + P_{drop})/P_{in}$ response, to reach the same amount of total optical power FS, the required electrical signal FS of Dual-MRMs is less than that of the notch-filter MRM. In other words, the dual-MRM architecture offers both E/O linearity and conversion-gain improvements by taking advantage of the $MRM_2$ Drop-Port peaking spectrum and doubled laser injection power (i.e., $2 \cdot P_{in}$) as shown in Fig. 2(d) and Fig. 1(a), respectively.

## III. Simulations and Rapid Experimental Validations

The linearity enhancement technique is simulated by using the O-band MRMs with less than 30-µm × 30-µm silicon area per footprint in GF45SPCLO. As shown in Fig. 3(a), the Dual-MRMs power-gain response is constructed by the linear summation of the $MRM_1$ and $MRM_2$ power-gain responses. By defining the normalized power-gain FS range from 0.25 to 0.75, which is equivalent to a 4.8-dB ER, the static E/O-conversion transfer curves of $MRM_1$ and Dual-MRMs w.r.t their normalized electrical signals at $V_P$ are shown in Fig. 3(b), where the Dual-MRMs transfer curve behaves as a way better straight line than the $MRM_1$ transfer curve does. Their deviations from an ideal case are quantified by the continuous INLs as shown in Fig. 3(c), where the peak-to-peak INL ($INL_{PP}$) of the Dual-MRMs transfer curve shows 2.65-bit (= 16.0 dB = ΔINL) static linearity improvement over the $MRM_1$ transfer curve.

The dynamic linearity for the high-frequency analog E/O conversions is evaluated by the IMDs in the $IP_3$ measurements with an 8-GHz analog RF signal at $V_P$. As shown in Fig 3(d), the IMD of Dual-MRMs exhibits 6.1-dB (= $\Delta SFDR_{IMD}$) linearity improvement over the IMD of $MRM_1$. Meanwhile, the dynamic linearity for the high-speed digital E/O conversions is evaluated by the differential nonlinearity (DNL) and INL of the optical eye diagrams with 40-Gb/s PAM-8 random data (i.e., 13.3 GSym/s and 3 bit/Sym data with 6.7-GHz Nyquist frequency) at $V_P$. As shown in Fig. 3(e) and 3(f), $DNL_{PP}$ and $INL_{PP}$ of the Dual-MRMs eye perform 2.36-bit (= 14.2 dB = ΔDNL) and 2.44-bit (= 14.7 dB = ΔINL) improvements,

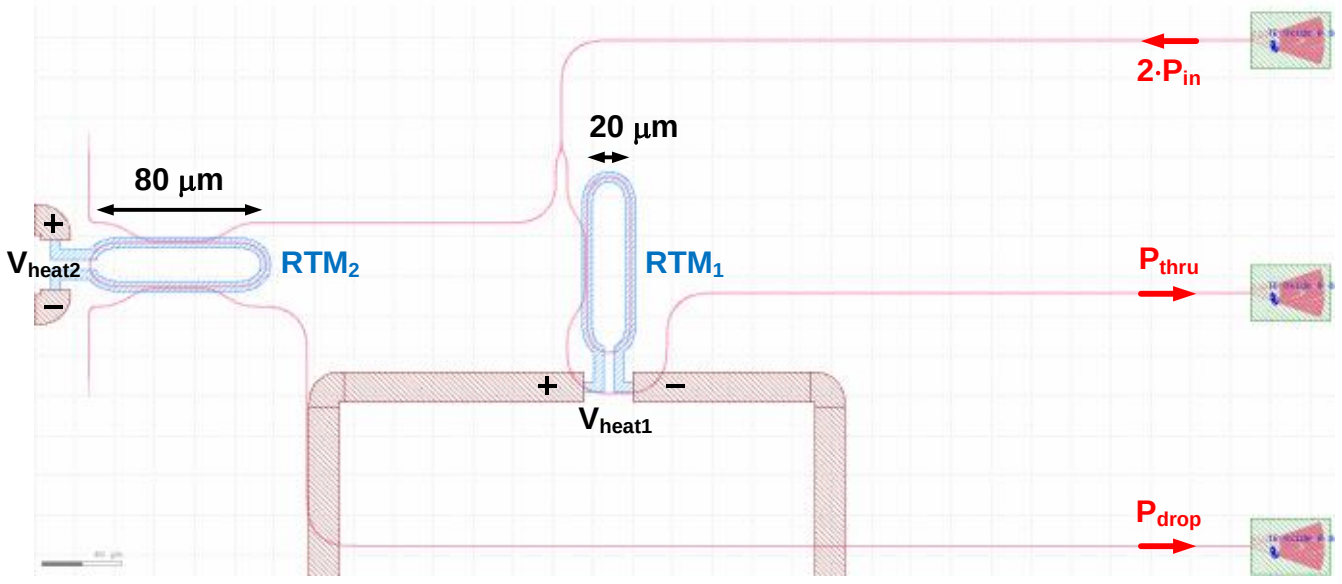

Fig. 4. The mask layout of the passive RTM linearity enhancement technique (Dual-RTMs: $RTM_1$ + $RTM_2$) for the rapid experimental validations.

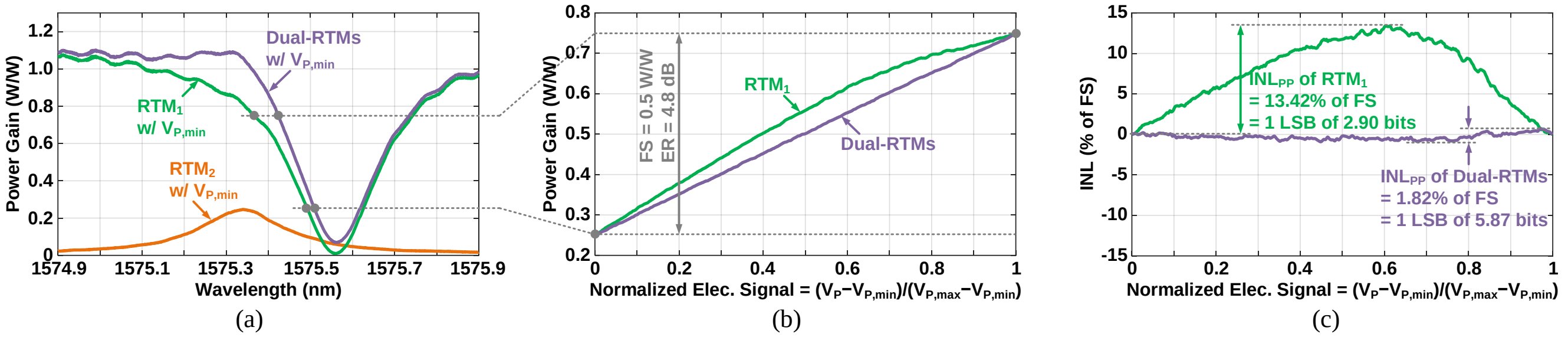

Fig. 5. The rapid experimental results by using C-band passive RTMs fabricated by Applied Nanotools Inc.: (a) The power-gain responses of $RTM_1$, $RTM_2$, and Dual-RTMs when $V_P = V_{P,min}$. (b) The static E/O-conversion transfer curves of $RTM_1$ and Dual-RTMs. (c) The INLs of the static E/O-conversion transfer curves in (b).

respectively, over the $MRM_1$ eye, within the 4.8-dB ER FS of the PAM-8 signal.

The initial test chip of the linearity enhancement technique has been rapidly fabricated by Applied Nanotools Inc. in 220-nm silicon-on-insulator (SOI) SiPh process technology for fast experimental validations in the lab. Instead of MRMs, the notch filter and bandpass filter are implemented by the racetrack modulators (RTM) as the mask layouts shown in Fig. 4, which can offer better flexibility in designing the coupling coefficients and corresponding quality factors as well as the peaking behaviors of the resonators without affecting the whole concept of the linearity enhancement. Though this rapid fabrication can only produce the passive RTMs while retaining their independent thermal control capabilities, the test chip can sufficiently demonstrate the power-gain responses, static E/O-conversion transfer curves, and measurable INLs for proof of concept as shown in Fig. 5(a), 5(b), and 5(c), respectively, with a tunable laser source (Keysight N7776C) and a dual-channel optical sampling scope (Keysight DCA-X). Under the pre-defined 4.8-dB ER FS, $INL_{PP}$ of the Dual-RTMs transfer curve shows nearly 3-bit ($\approx$ 17.9 dB = $\Delta$INL) static linearity improvement over the $RTM_1$ transfer curve.

## IV. Discussions and Future Work

It is important to note that the summation process of $P_{thru}$ and $P_{drop}$ in Fig. 1(a) is done by the intensity detections of a dual-input photodiode (PD) during the O/E conversion at the far-end receiver (RX) side to generate the photocurrent $I_{PD} = (I_{thru} + I_{drop})$, which is linearly proportional to $(P_{thru} + P_{drop})$ based on the PD responsivity in units of A/W, and then the following trans-impedance amplifier (TIA) linearly converts $(I_{thru} + I_{drop})$ to $(V_{thru} + V_{drop})$ in the electronic domain. In reality, not only the non-idealities of PD and TIA could further degrade the overall optical link linearity, but also the latency mismatch between the two fibers could cause the propagations of the power envelops, i.e., $P_{thru}(t)$ and $P_{drop}(t)$, not to reach PD at the same moment of time, which is the main practical concern of this linearity enhancement technique. For the example of an 8-GHz RF signal at $V_P$, a 10-ps latency mismatch between the two 10-km long fibers can induce a certain amount of signal distortion when the $I_{thru}(t)$ and $I_{drop}(t)$ waveforms are summed in the time domain with an 8% (= 10-ps/125-ps) phase error. This issue can be resolved by the modified dual-MRM architecture shown in Fig. 1(b) having a single fiber and two laser injections with well-spaced wavelengths, i.e., $\lambda_1$-laser and $\lambda_2$-laser, so that $P_{thru}(t)$ and $P_{drop}(t)$ carried by $\lambda_1$-laser and $\lambda_2$-laser, respectively, can be both coupled to a single fiber at the near-end TX side to maintain an identical fiber latency without interfering with each other. This modified dual-MRM architecture doesn't change the concept and analysis of the power-gain responses shown in Fig. 2, but the $P_{thru}/P_{in}$ and $P_{drop}/P_{in}$ responses for Fig. 1(b) shall have an extra static wavelength offset because of the delta between $\lambda_1$ and $\lambda_2$, which seems to cause an improper $(P_{thru} + P_{drop})/P_{in}$ response in the frequency (or wavelength) domain. However, the linearity enhancement is still valid because the linear summation actually happens in the time-domain power envelopes, i.e., $P_{thru}(t)$ of $\lambda_1$-laser and $P_{drop}(t)$ of $\lambda_2$-laser, not in the frequency domain.

The dual-MRM architecture in Fig. 1(b), including all electronic circuits, photonic devices, and MRMs with PN-junction reverse-bias modulation and thermal control capabilities, will be monolithically integrated and fabricated in GF45SPCLO to demonstrate a fully on-chip energy/area-efficient E/O interface with a high degree of linearity for various analog and digital optical communication systems.